\documentstyle[12pt]{article}

\begin{document}

\setcounter{page}{1} \pagestyle{plain} \vspace{1cm}
\begin{center}
\Large{\bf Phantom-Like Behavior in $f(R)$-Gravity}\\
\small \vspace{1cm} {\bf Kourosh
Nozari\footnote{knozari@umz.ac.ir}}\quad and \quad
{\bf Tahereh Azizi\footnote{t.azizi@umz.ac.ir}}\quad  \\
\vspace{0.5cm} {\it Department of Physics,
Faculty of Basic Sciences,\\
University of Mazandaran,\\
P. O. Box 47416-1467, Babolsar, IRAN}\\
\end{center}
\vspace{1.5cm}
\begin{abstract}
We investigate possible realization of the phantom-like behavior in
the framework of $f(R)$-gravity models where there are no phantom
fields in the matter sector of the theory. By adopting some
observationally reliable ansatz for $f(R)$, we show that it is
possible to realize phantom-like behavior in $f(R)$-gravity without
introduction of phantom fields that suffer from instabilities and
violation of the null energy condition. Depending on the choice of
$f(R)$, the null energy condition is fulfilled in some subspaces of
each model parameter space.

\end{abstract}
\newpage
\section{Introduction}
One of the most astonishing discoveries of the last two decades is
the observation of a positively accelerated phase of cosmic
expansion. This amazing result comes from several sources of
observational data such as: measurements of luminosity-distances of
supernovae type Ia (SNIa) [1], the cosmic microwave background (CMB)
temperature anisotropies with the Wilkinson Microwave Anisotropy
probe (WMAP) satellite [2], large scale structure [3], the
integrated Sachs-Wolfe effect [4], and the weak lensing [5]. Indeed,
general relativity with ordinary matter content of the universe
leads to a decelerating universe and therefore it cannot describe
this accelerating expansion which has been confirmed by a huge
amounts of observational data. In order to realize this late-time
acceleration theoretically, several approaches have been proposed.
One possibility is to consider an extra source of energy-momentum
with a negative pressure in the matter sector of the Einstein field
equations. However, the nature of this extra component (the so
called dark energy) is yet unknown for cosmologists. A very simple
and popular candidate for dark energy proposal is the cosmological
constant [6], but this scenario suffers from some serious problems
such as huge amount of fine-tuning and coincidence problems. Beside
these problems, this scenario has not a dynamical behavior because
of a constant equation of state parameter ($\omega_{\Lambda}=-1$).
Another suggestion for dark energy is the dynamical models that
include various scalar fields such as quintessence, k-essence,
chaplygin gas, phantom fields, quintom fields and so on [7]. On the
other hand, one of the most important results of the observational
data comes from WMAP5 that the equation of state parameter of dark
energy can be less than $-1$ and even can have a transient behavior
[8]. While general relativity with one scalar field cannot realize
such a crossing behavior, non-minimal coupling of scalar field and
gravity leads to this crossing phenomenon [9].

There is another approach to realize the cosmic speedup: modifying
geometric part of the gravitational theory. This proposal can be
realized in braneworld scenario ( DGP model and its extensions
[10]), string inspired scenarios ( Gauss-Bonnet terms in the action
[11]) and so on. A very popular modified gravity model is the so
called $f(R)$-gravity [12] where $f(R)$ is an arbitrary function of
the scalar curvature $R$. This scenario has the interesting feature
that choosing an observationally reliable $f(R)$, it is possible to
describe the early inflation as well as the late time acceleration
of the universe in a fascinating manner [12]. Recently it has been
shown that one can realize the phantom-like effect ( increasing of
the effective dark energy density with cosmic time and an equation
of state parameter less than $-1$) in the normal branch of the DGP
cosmological solution without introducing any phantom fields that
violate the null energy condition (NEC) [13,14]. This type of the
analysis then has been extended by several authors [15]. The main
goal in these studies is the realization of the phantom-like
behavior without introducing any phantom fields in the matter sector
of the theory. In fact, since phantom fields suffer from
instabilities and violate the null energy condition, it is desirable
to realize this behavior without introduction of phantom fields.
With this motivation, in this paper we introduce another alternative
to realize phantom like effect: We study possible realization of
this behavior in the framework of $f(R)$-gravity models. We consider
some observationally reliable versions of $f(R)$ gravity and
investigate the phantom-like behavior of each model without
introducing any phantom field that violates the null energy
condition. Some of these model such as Hu and Sawicki (HS) model
have passed the solar system tests in a very good manner as well as
the perturbation theory [16]. We show that all of these models in
some subspaces of the model parameter space realize a phantom-like
behavior without introducing any phantom fields. We study the
conditions that are required in each case to fulfill the null energy
condition.

\section{$f(R)$-gravity }
In this section we consider the metric formalism of $f(R)$-gravity
and we summarize the field equations of the scenario. The action of
a general $f(R)$-gravity theory is given by [12,17,18,19]
\begin{equation}
S=\frac{1}{2\kappa}\int d^{4}x\sqrt{-g}\{f(R)+{\cal{L}}_{M}\},
\end{equation}
where $R$ is the scalar curvature, $f(R)$ is an arbitrary function
of $R$ and $\kappa=8\pi G$ is the gravitational constant. The term
${\cal{L}}_{M}$ accounts for the matter content of the universe.
Using the metric approach, variation of this action with respect to
$g_{\mu\nu}$ provides the field equation
\begin{equation}
f'(R)R_{\mu\nu}-\frac{1}{2}f(R)g_{\mu\nu}-\nabla_{\mu}\nabla_{\nu}f'(R)+g_{\mu\nu}\Box
f'(R)=\kappa T_{\mu\nu}^{(M)},
\end{equation}
where the prime denotes derivative with respect to $R$ and the
matter stress-energy density is defined as
\begin{equation}
T_{\mu\nu}^{(M)}=-\frac{2}{\sqrt{-g}}\frac{\delta(\sqrt{-g}{\cal{L}}_{M})}{\delta
(g^{\mu\nu})}.
\end{equation}
By assuming a spatially flat FRW metric, the Friedmann equation can
be written as
\begin{equation}
H^{2}=\frac{8\pi G}{3}\Big[\frac{\rho_{M}}{f'(R)}+\rho_{curv}\Big],
\end{equation}
where $\rho_{M}$ is the energy density of the ordinary matter and
$\rho_{curv}$ is the energy density of the \emph{curvature fluid}
defined as
\begin{equation}
\rho_{curv}=\frac{1}{f'(R)}\bigg\{\frac{1}{2}[f(R)-Rf'(R)]-3H\dot{R}f''(R)\bigg\}.
\end{equation}
Throughout this paper we consider the Jordan frame, thus the
continuity equation of the matter sector can be read as usual
\begin{equation}
\rho_{M}=\rho_{M}(t=t_{0})=3H_{0}^{2}\Omega_{M}(1+z)^{3}.
\end{equation}
Where $\Omega_{M}$ is the present day matter density parameter. The
continuity equation for the curvature fluid is given in the
following form [17]
\begin{equation}
\dot{\rho}_{curv}+3H(1+\omega_{curv}\rho_{curv})=\frac{3H_{0}^{2}
\Omega_{M}\dot{R}f''(R)(1+z)^{3}}{\big[f'(R)\big]^{2}}.
\end{equation}
By definition, the pressure of the curvature fluid is given by
\begin{equation}
P_{curv}=\frac{1}{f'(R)}\bigg\{2H\dot{R}f''(R)+\ddot{R}f''(R)+
\dot{R}^{2}f'''(R)-\frac{1}{2}[f(R)-Rf'(R)]\bigg\}.
\end{equation}
The equation of state parameter corresponding to the curvature
sector of the theory can be read as follows
\begin{equation}
\omega_{curv}=-1+\frac{\ddot{R}f''(R)+\dot{R}\Big[\dot{R}f'''(R)-
Hf''(R)\Big]}{\frac{1}{2}[f(R)-Rf'(R)]-3H\dot{R}f''(R)}.
\end{equation}
From the continuity equations (7) and field equation (4), the Hubble
rate can be expressed as follows
\begin{equation}
\dot{H}=-\frac{1}{2f'(R)}\bigg\{3H_{0}^{2}\Omega_{M}(1+z)^{3}+\ddot{R}f''(R)+
\dot{R}\Big[\dot{R}f'''(R)-Hf''(R)\Big]\bigg\},
\end{equation}
where $R=6(\dot{H}+2H^{2})$. This equation is a very complicated
equation and it is very difficult to solve it even with the simplest
forms of the $f(R)$-gravity.
\section{ Phantom-like behavior of $f(R)$-gravity }
With phantom-like behavior, we mean an effective energy density
which is positive and grows with time and its equation of state
parameter stays less than $-1$. In this section, by adopting some
cosmologically viable ansatz, we show that the modified gravity can
lead to the effective phantom dark energy and phantom-like behavior
without need to introduce any kind of the phantom (negative energy
density) scalar fields that violate the null energy condition ( see
also [18], [19] for some earlier attempts in this regard). To do
this end, the modified Friedmann equation (4) can be expressed in a
familiar form
\begin{equation}
H^{2}=\frac{8\pi G_{eff}}{3}\Big[\rho_{M}+f'(R)\rho_{curv}\Big].
\end{equation}
This relation shows that in $f(R)$ gravity the gravitational
constant $G$ can be replaced by an effective gravitational constant
$G_{eff}=\frac{G}{f'(R)}$. The equation (11) can be recast in the
following form
\begin{equation}
H^{2}=\frac{8\pi G_{eff}}{3}\Big[\rho_{M}+\rho_{eff}\Big],
\end{equation}
where $\rho_{eff}=f'(R)\rho_{curv}$ with  $\rho_{curv}$ defined as
(5) and therefore effective equation of state parameter is given by
\begin{equation}
\omega_{eff}=-\frac{1}{f'(R)}+\frac{\ddot{R}f''(R)+\dot{R}\Big[\dot{R}f'''(R)-Hf''(R)\Big]}
{f'(R)\bigg(\frac{1}{2}[f(R)-Rf'(R)]-3H\dot{R}f''(R)\bigg)}.
\end{equation}
Now we have all necessary ingredients to discuss phantom-like
behavior of $f(R)$-gravity. To do this end, we consider some
observationally reliable ansatz for $f(R)$.

\subsection{Phantom-like effect with $f(R)=R+f_{0}R^{\alpha}$}
Phantom-like behavior is the growth of the effective energy density
with cosmic time and in the same time, the effective equation of
state parameter should stay always less than $-1$. we mean as a
first illustrative example, we consider the following ansatz [19]
\begin{equation}
f(R)=R+f_{0}R^{\alpha}
\end{equation}
with constant $f_{0}$ and $\alpha$. If $\alpha<1$, in the small
curvature limit the second term dominates. Note that in this class
of models, a negative $\alpha$ implies the presence of a term
inversely proportional to $R$ in the action that can lead to the
present cosmic speed-up [20]. For $\alpha=0$ the curvature freezes
into a fixed value lead to producing a class of models that
accelerate in a manner similar to the cosmological constant included
models ($\Lambda CDM$). As has been stated, for $\alpha=-1$ this
model can describe the acceleration of the universe, but the model
evolves in the future into an unstable regime where $1+f'(R)<0$ and
it does not contain $\Lambda CDM$ as a limiting case of parameter
space [21]. Now the expression for effective quantities in this
model take the following form
\begin{equation}
\rho_{eff}=(1-\alpha)f_{0}R^{\alpha}\Big[\frac{1}{2}+\frac{3\alpha
H\dot{R}}{R^2}\Big]
\end{equation}
 and
\begin{equation}
\omega_{eff}=\frac{-1}{1+\alpha
f_{0}R^{\alpha-1}}\bigg(1+\frac{\ddot{R}+\frac{\dot{R}^2}{R}(\alpha-2)-
H\dot{R}}{\frac{R^2}{2\alpha}+3H\dot{R}}\bigg).
\end{equation}
To have an intuition of phantom-like behavior in this case, we adopt
the ansatz $a(t)= a_{0}t^{\nu}$. It is important to note that this
is a solution of the Friedmann equation in our case. Especially, for
$\nu > 1$ it gives an accelerating universe which is essentially
realizable in $f(R)$-gravity [12]. In table $1$ we have shown the
acceptable ranges of $\alpha$ to realize phantom-like behavior in
some subsets of the model parameter space. As mentioned before,
theoretically negative values of $\alpha$ can account for cosmic
acceleration, but they evolve in an unstable regime in the future.
Especially, it is clear that for $\alpha=-1$ the null energy
condition is violated. While for negative values of $\alpha$,
$\rho_{eff}$ grows with decreasing $z$ but its values always remain
negative and the effective equation of state parameter is
quintessence-like. The case $\alpha=0$ is corresponding to an
effective cosmological constant with equation of state parameter
$\omega_{eff}=-1$. Our numerical analysis shows that in this case
phantom-like behavior can be realized if $0.5\leq\alpha<1$. The case
$\alpha=1$ with a redefinition of the Newtonian gravitational
constant is corresponding to general relativity. For $\alpha>1$, the
effective equation of state parameter lies in the non-phantom region
of parameter space and therefore it cannot account for phantom-like
behavior. In figure $1$ we plot the effective energy density and
equation of state parameter of the model versus the redshift $z$ for
$\alpha=0.5$ (note that this choice is corresponding to
$f(R)=R+f_{0}\sqrt{R}$ model). As figure shows, in this case the
effective energy density increases with decreasing $z$ and the
effective equation of state parameter is less than $-1$ a typical
realization of the phantom-like effect. From figure $2$ we can
derive the acceptable ranges of $\alpha$ to fulfill the null energy
condition. As this figure shows, null energy condition is respected
in some subspaces of the model parameter space and not in the entire
parameter space. we note that for $\alpha<0$ the phantom-like
prescription breaks down since in this case $\rho_{eff}$ is
negative.
\begin{table}
\begin{tiny}
\begin{center}
\caption{Acceptable range of $\alpha$ to have a phantom-like
behavior with $f(R)=R+f_{0}R^{\alpha}$. } \vspace{0.5 cm}
\begin{tabular}{|c|c|c|c|c|c|c|c|c|c|}
  \hline
  \hline $\alpha$ &$\alpha<-2$ & $-2\leq\alpha<-1.5$& $-1.5<\alpha\leq-1$&$-1\leq\alpha<-0.5$&$-0.5\leq\alpha<0$&$\alpha=0$&$0<\alpha<0.5$&$0.5\leq\alpha<1$&$\alpha\geq1 $\\
  \hline $\rho_{eff}$&negative& negative &negative& negative &negative&cosmological constant&positive&positive&positive \\
  \hline $\omega_{eff}$&$\omega_{eff}>-1$ &$\omega_{eff}>-1$&$\omega_{eff}>-1$ &$\omega_{eff}>-1$&$\omega_{eff}>-1$& $\omega_{eff}=-1$&$\omega_{eff}<-1$&$\omega_{eff}<-1$&$\omega_{eff}>-1$\\
 \hline NEC &not respected  & respected & not respected & respected& respected & respected & not respected & respected&respected \\
   \hline
\end{tabular}
\end{center}
\end{tiny}
\end{table}

\begin{figure}[htp]
\begin{center}
\includegraphics{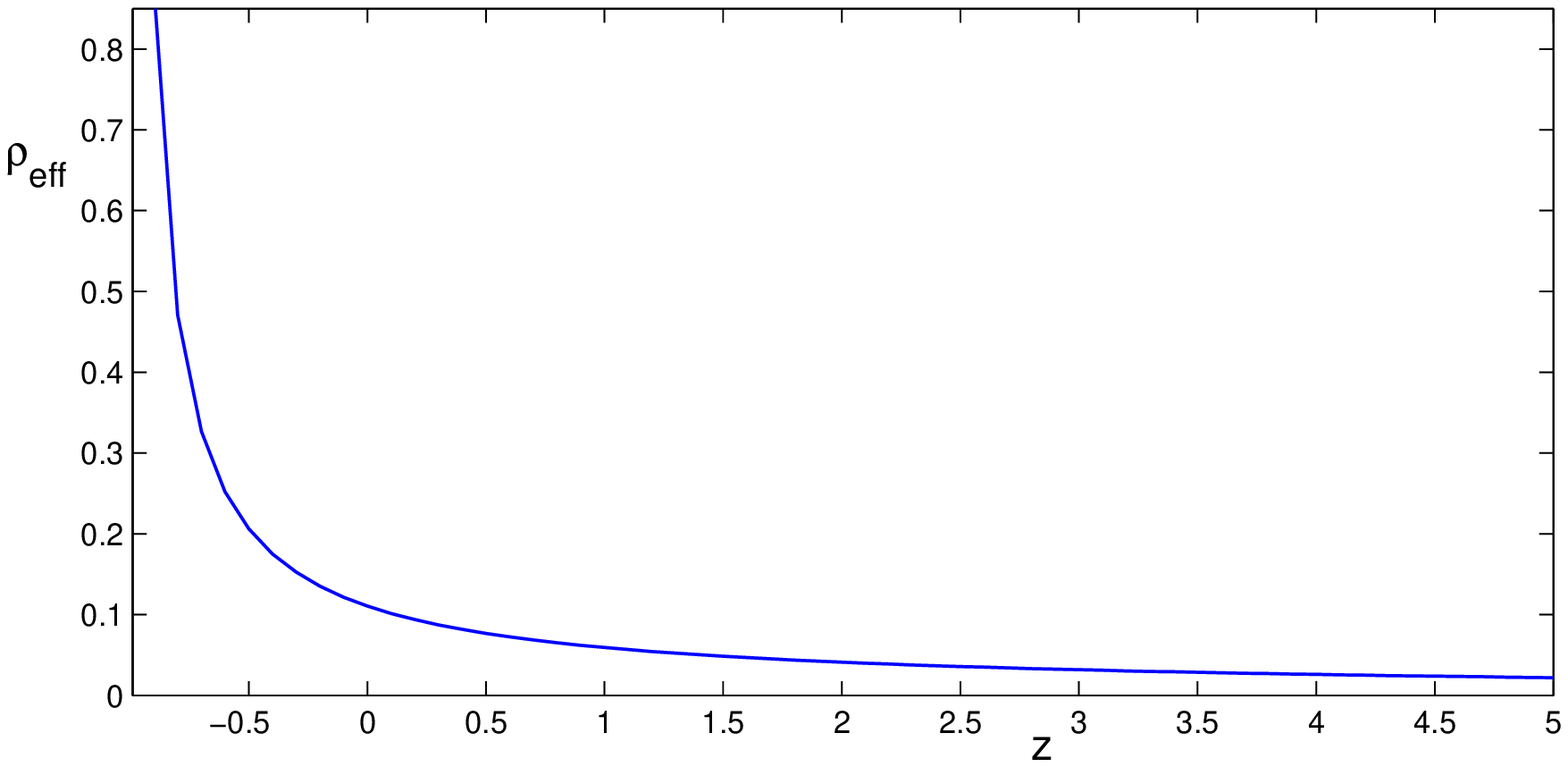} \vspace{1cm}\includegraphics{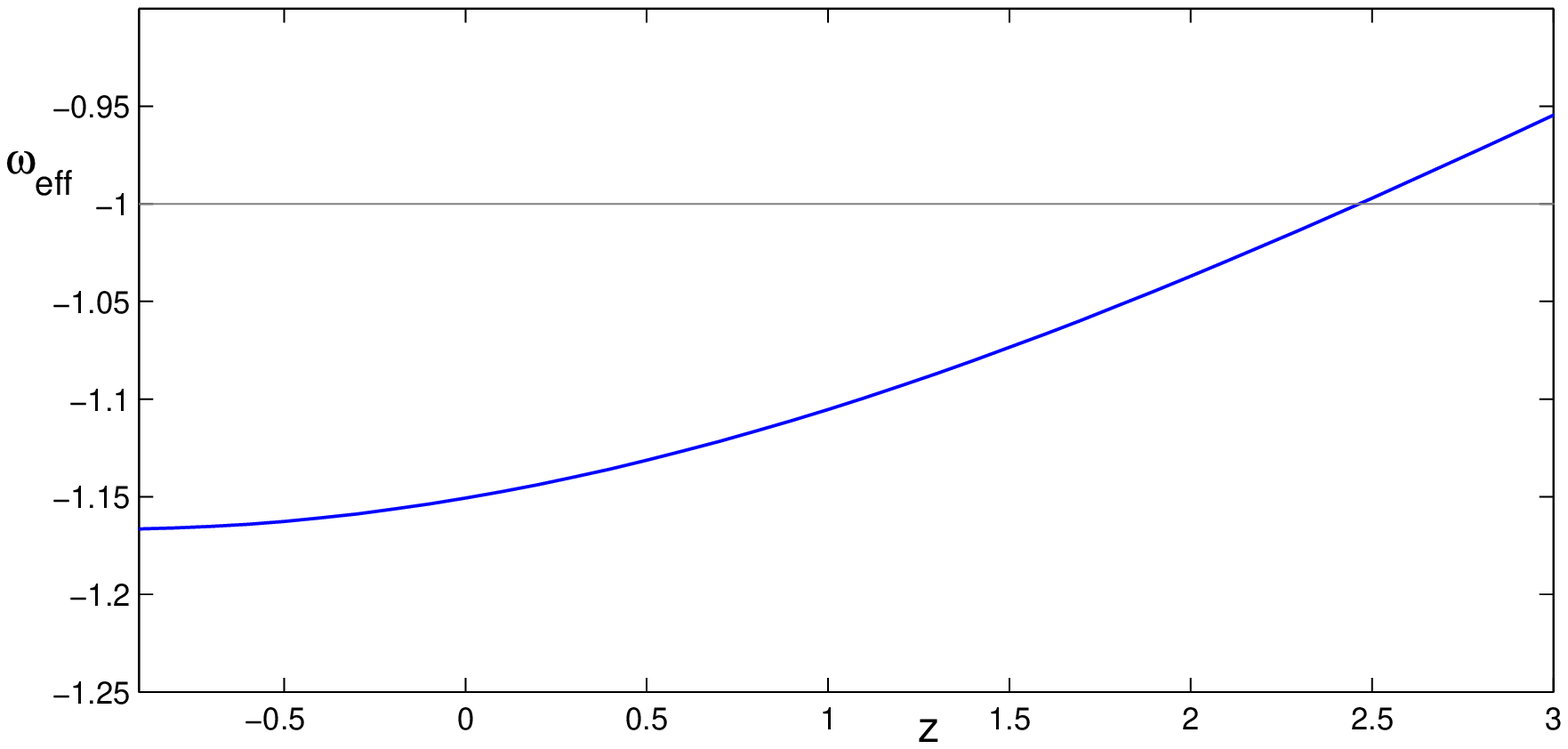}\vspace{5cm}
\end{center}
\vspace{1cm} \caption{\small {Variation of the effective dark energy
density versus the redshift (left hand side). The effective dark
energy density increases with decreasing $z$ and therefore shows a
phantom-like behavior. The effective equation of state parameter
versus redshift (right hand side) which has entered in the phantom
phase in the past.}}
\end{figure}

\begin{figure}[htp]
\begin{center}\includegraphics{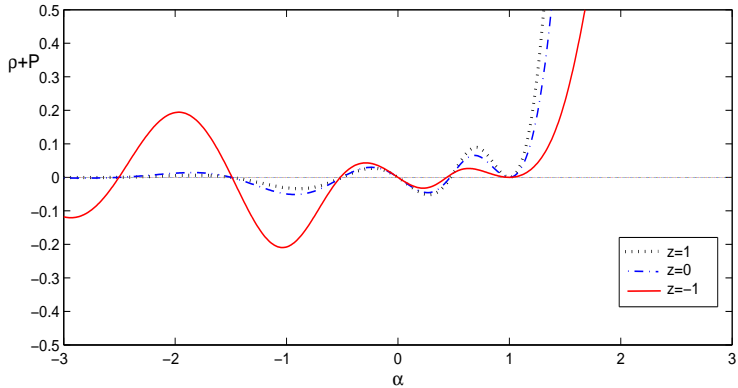} \vspace{4cm}
\end{center}
 \caption{\small {$\rho_{eff}+p_{eff}$ versus $\alpha$ with $z=\pm1,0$.
  The null energy condition is violated for some values of $\alpha$, but
  there are subspaces of the model parameter space that respect this condition.}}
\end{figure}

\subsection{Phantom-like effect with $\ln R$ term }
In this subsection we consider a modified gravity scenario with $\ln
R$ term in the form [19]
\begin{equation}
f(R)=R+\beta \ln\frac{R}{\mu^{2}}+\gamma R^{m}.
\end{equation}
The second term in this ansatz containing $\ln R$, is growing at
small curvature. Basically this term is induced by quantum effects
in curved spacetime. It has been shown that this model has a well
defined Newtonian limit and is able to provide the late time
acceleration without need to introduce any dark energy component
[19]. Choosing $m=2$, this model leads to a very interesting result:
unification of the early time inflation and the late time
acceleration. On the other hand, considering $R^{2}$ term can
suppress the instabilities arises in the perturbation theory of the
model as well as improving the solar system bounds, consequently the
theory can be viable [19].

Now, the effective quantities in this model attain the following
forms
\begin{equation}
\rho_{eff}=-\frac{\beta}{2\mu^2}+\frac{\beta}{2}\ln\frac{R}{\mu^{2}}+(1-m)\gamma
R^2\Big(\frac{1}{2}+\frac{3mH\dot{R}}{R^2}\Big)+\frac{3\beta
H\dot{R}}{\mu^{2}R^2}
\end{equation}
and
$$\omega_{eff}=\frac{-1}{1+\frac{\beta}{\mu^{2}R}+m\gamma R^{m-1}}$$
\begin{equation}
+ \frac{(\ddot{R}-H\dot{R})\Big[\frac{-\beta}{\mu^2R^2}+m(m-1)\gamma
R^{m-2}\Big]+ \frac{2\beta}{\mu^2R}+m(m-1)(m-2)\gamma
R^{m-1}}{\Big(1+\frac{\beta}{\mu^{2}R}+m\gamma
R^{m-1}\Big)\Big[-\frac{\beta}{2\mu^2}+\frac{\beta}{2}\ln\frac{R}{\mu^{2}}+(1-m)\gamma
R^2\Big(\frac{1}{2}+\frac{3mH\dot{R}}{R^2}\Big)+\frac{3\beta
H\dot{R}}{\mu^{2}R^2}\Big]}.
\end{equation}
In figure $3$ we plot the effective energy density and equation of
state parameter versus the redshift $z$ for $m=2$. As this figure
shows, in this case the effective energy density has a growing
behavior with decreasing $z$, so it displays the phantom-like
behavior without introducing any phantom field. The effective
equation of state parameter in the late times lies in the phantom
region with no crossing behavior. To realize phantom divide line
crossing we can introduce for instance a canonical scalar field in
the matter sector of the theory. In figure $4$ we have investigated
the acceptable ranges of $m$ to satisfying the null energy condition
for $z=\pm 0.2,\,0$. This figure shows that this model respects the
null energy condition for $m\geq1.3$. As has been pointed out in
[19], the presence of higher derivative terms like $R^{2}$ (which
may be responsible for early time inflation) in this model helps one
to pass the existing arguments such as instabilities and solar
system tests against such modification of the Einstein gravity.
\begin{figure}[htp]
\begin{center}
\includegraphics{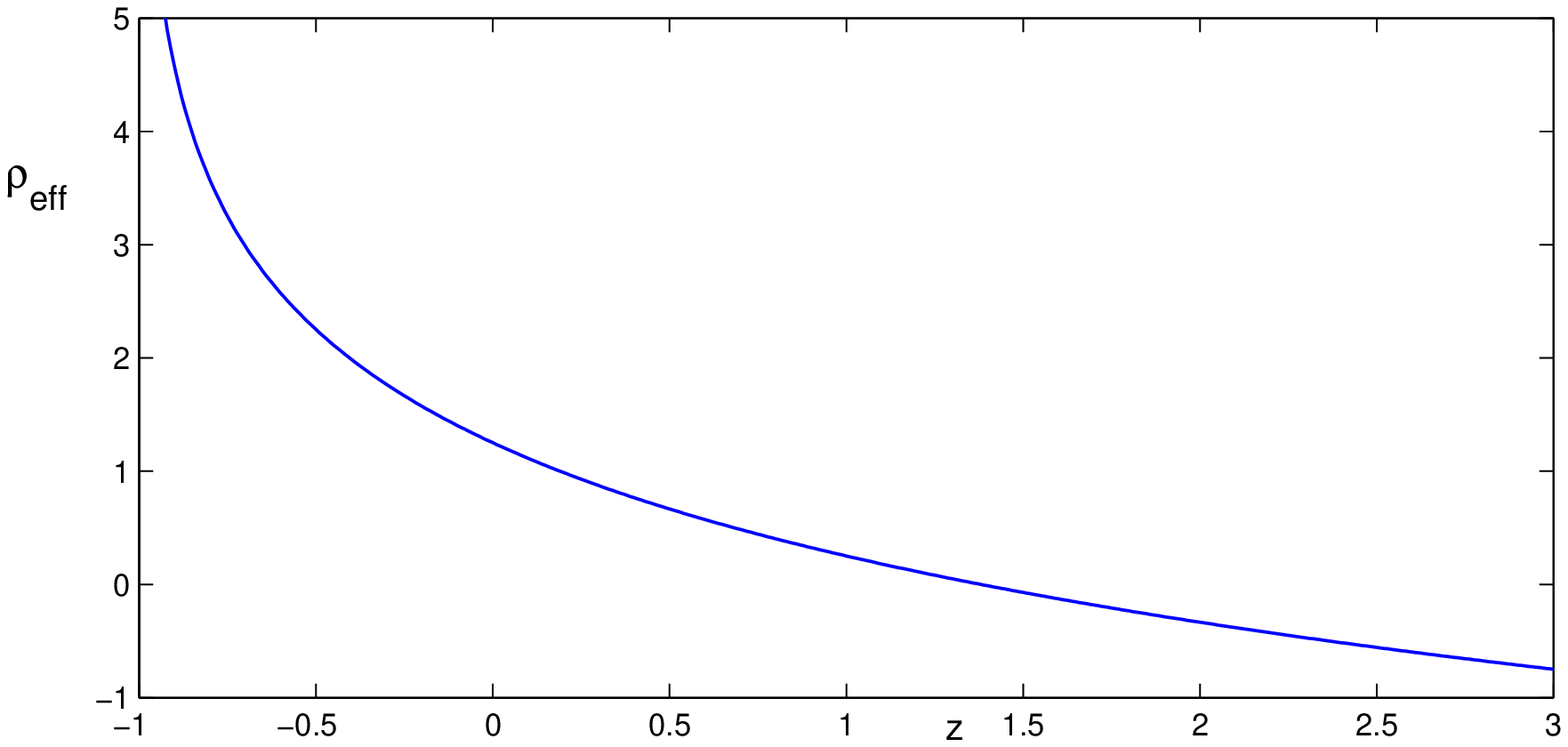} \vspace{1cm}\includegraphics{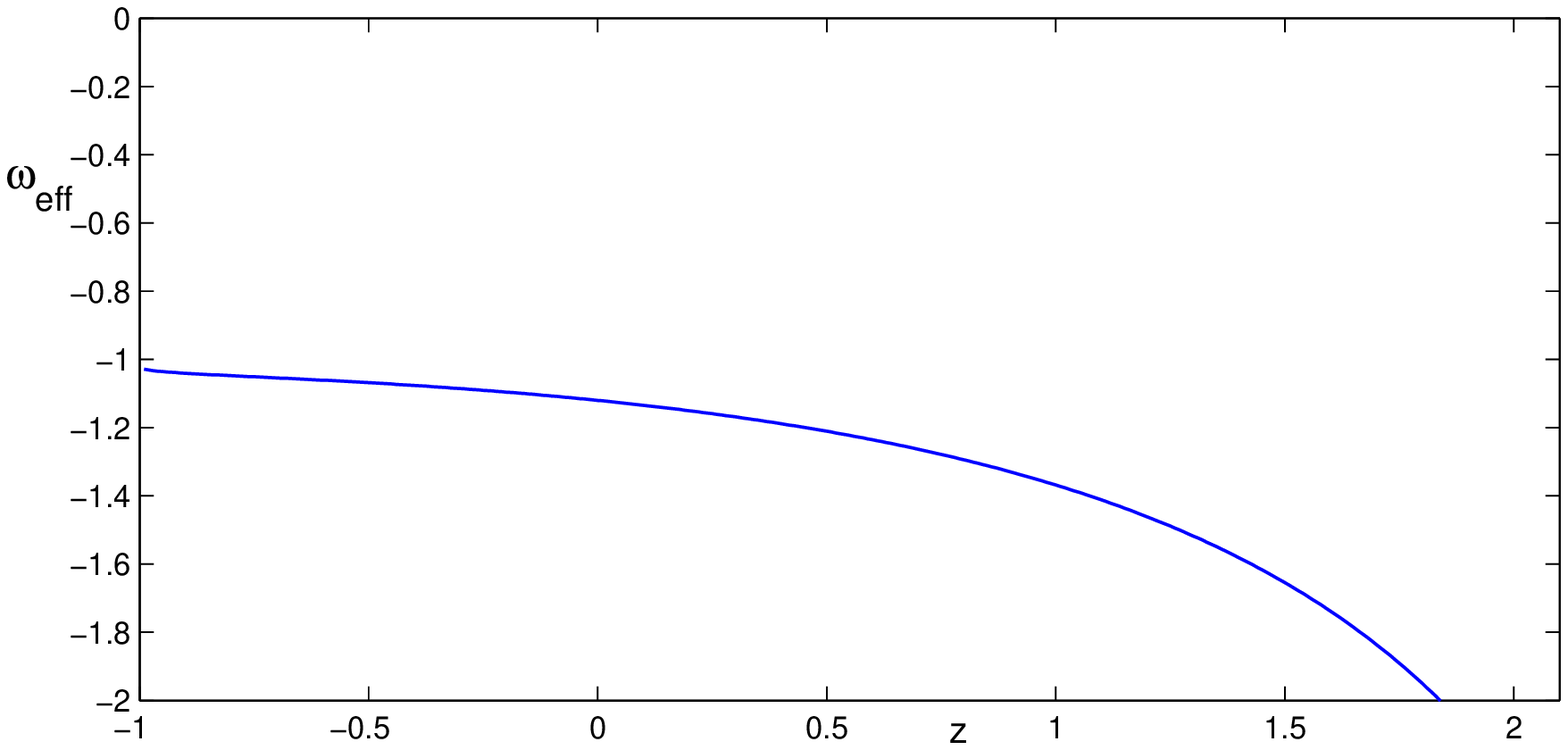}\vspace{2.5cm}
\end{center}
\vspace{1cm} \caption{\small { Variation of the effective dark
energy density versus the redshift (left hand side). The effective
dark energy density increases with decreasing $z$, so it shows a
phantom-like behavior. The effective equation of state parameter is
less than $-1$ in the small redshifts ( right hand side). }}
\end{figure}

\begin{figure}[htp]
\begin{center}\includegraphics{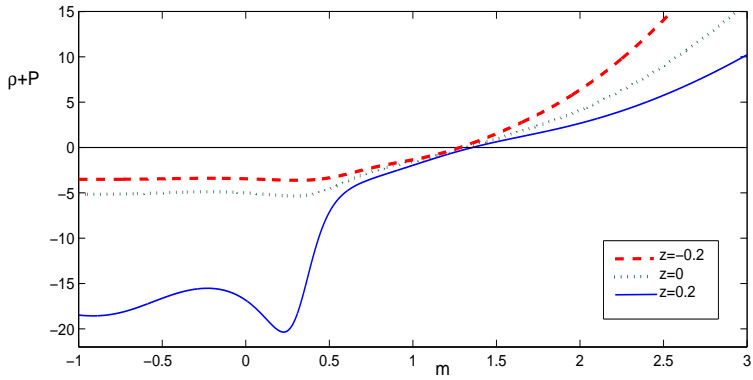} \vspace{4.5cm}
\end{center}
 \caption{\small {The null energy condition is fulfilled for $m\geq1.3$.
 This figure is plotted for the redshifts $z=\pm0.2,0$.
   }}
\end{figure}
\newpage

\subsection{The Hu-Sawicki model}
One of the most interesting modified gravity model has been proposed
by Hu and Sawicki (HS model [22]) that can escape the severe
constraint imposed by the solar system tests. The form of $f(R)$ in
this model is written as follows
\begin{equation}
f(R)=R-M^{2}\frac{c_{1}\Big(\frac{R}{M^2}\Big)^n
}{c_{2}\Big(\frac{R}{M^2}\Big)^n+1},
\end{equation}
where $n>0$ and $c_{1}$ and $c_{2}$ are arbitrary dimensionless
constants while $M$ has the dimension of mass. This model yields an
effective cosmological constant which generates the late-time
accelerated expansion [13]. For $R\gg M^{2}$, equation (20) can be
expanded to find
\begin{equation}
f(R)\approx
R-M^{2}\frac{c_{1}}{c_{2}}+M^{2}\frac{c_{1}}{c_{2}^2}\Big(\frac{M^{2}}{R}\Big)^{n}.
\end{equation}
In the limit of $\frac{c_{1}}{c_{2}^{2}}\rightarrow0$ at fixed
$\frac{c_{1}}{c_{2}}$, this can be realized as an effective
cosmological constant $\Lambda_{eff}=M^{2}\frac{c_{1}}{c_{2}}$ which
produces the late time acceleration of the universe. The effective
energy density in this model is given as follows
\begin{equation}
\rho_{eff}=\frac{A}{2}+\frac{A}{2c_{2}(\frac{R}{M^2})^n+1}
+\frac{2Ac_{2}(\frac{R}{M^2})^n}{[c_{2}(\frac{R}{M^2})^n+1]^2}\bigg(1-\frac{6H\dot{R}}{R^2}\bigg)-
\frac{3H\dot{R}Ac_{2}n^{2}(\frac{R}{M^2})^n}{\Big[c_{2}(\frac{R}{M^2})^n+1\Big]^{3}R^2}
\Big[c_{2}(\frac{R}{M^2})^n-1\Big],
\end{equation}
where $A\equiv M^{2}\frac{c_{1}}{c_{2}}$. The effective equation of
state parameter is a lengthy expression and we do not write it here
explicitly. Figure $5$ ( left hand side) shows the behavior of the
effective energy density versus the redshift for $n=4$. Similar to
previous cases, the effective energy density increases with
decreasing $z$. The effective equation of state parameter is in the
phantom phase too, but it never crosses the phantom divide line (
figure $5$, right hand side). Figure $6$ shows the acceptable ranges
of $n$ to fulfill the null energy condition. Table $2$ shows the
appropriate subspaces of the model parameter space to have
phantom-like behavior and fulfilling the null energy condition in
the HS model.
\newpage
\begin{table}
\begin{tiny}
\begin{center}
\caption{Acceptable range of $n$ to have a phantom like behavior. }
\vspace{0.5 cm}
\begin{tabular}{|c|c|c|c|c|c|c|c}
  \hline
  \hline value of $n$ &$n<1.1$ & $1.1\leq n<1.7$& $1.7<n\leq2.2$&$2.2\leq n<2.7$&$2.7\leq n<3.4$&$n\geq3.4$  \\
  \hline $\rho_{eff}$&growing but negative& growing but negative&decreasing& decreasing &growing but negative&growing and positive \\
  \hline $\omega_{eff}$&$\omega_{eff}>-1$ &$\omega_{eff}>-1$&$\omega_{eff}<-1$ &$\omega_{eff}<-1$&$\omega_{eff}>-1$& $\omega_{eff}<-1$\\
 \hline null energy condition & not respected & respected & not respected & respected& not respected & respected  \\
   \hline
\end{tabular}
\end{center}
\end{tiny}
\end{table}

\begin{figure}[htp]
\begin{center}
\includegraphics{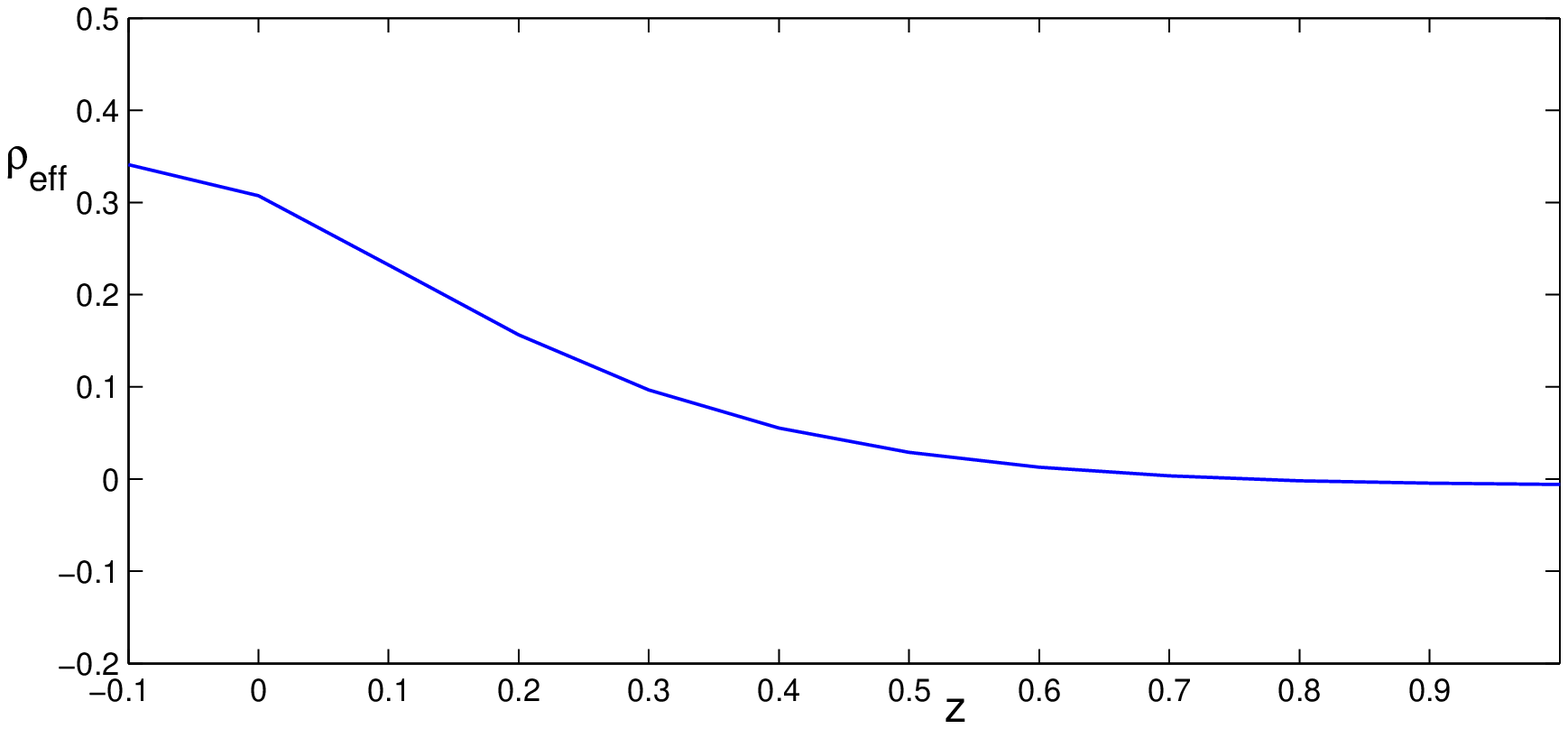} \vspace{1cm}\includegraphics{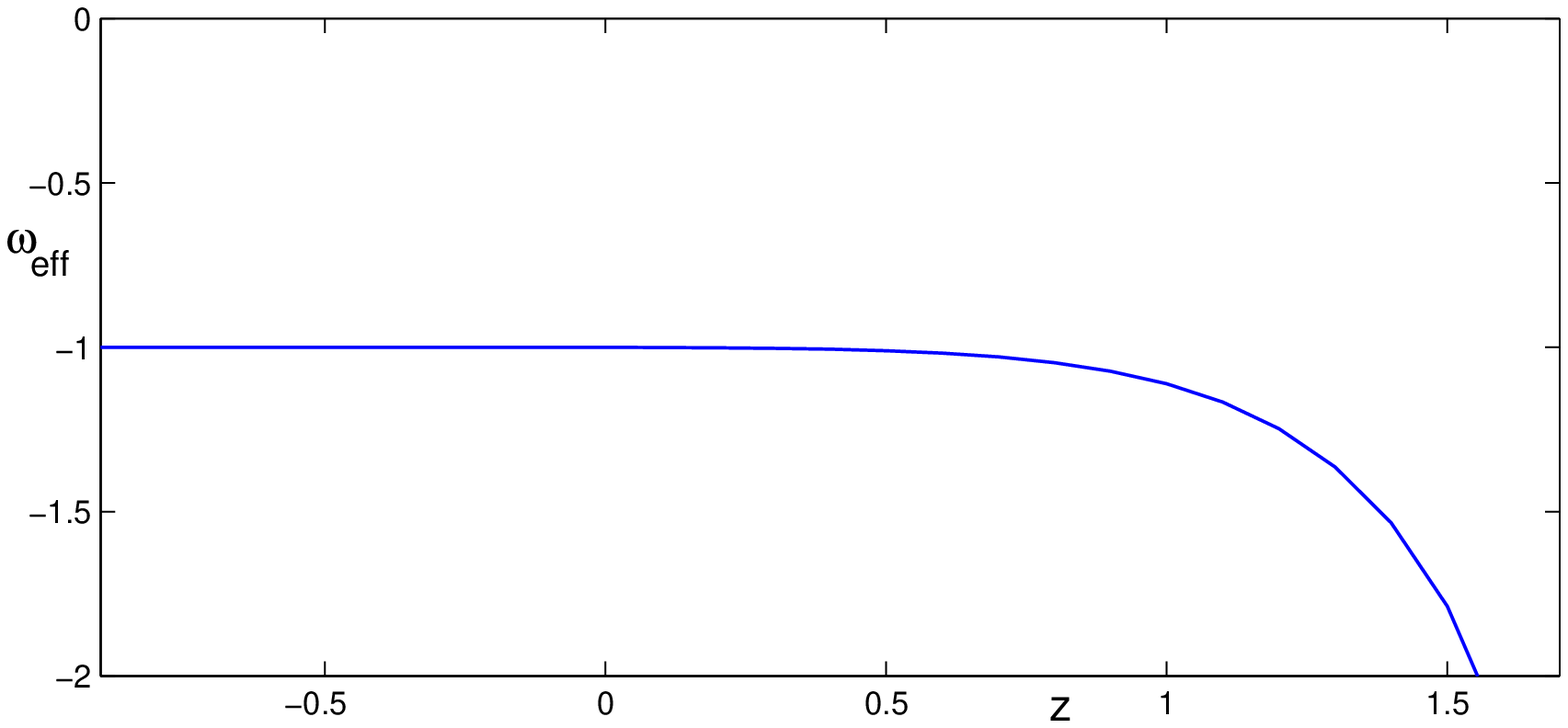}\vspace{4cm}
\end{center}
\vspace{2cm} \caption{\small { The effective dark energy density
versus the redshift (left hand side). The effective dark energy
density increases with decreasing $z$ and therefore shows a
phantom-like behavior. The effective equation of state parameter
remains in the phantom region for small values of redshift (present
day and future times evolution of the universe)(right hand side).}}
\end{figure}
\begin{figure}[htp]
\begin{center}\includegraphics{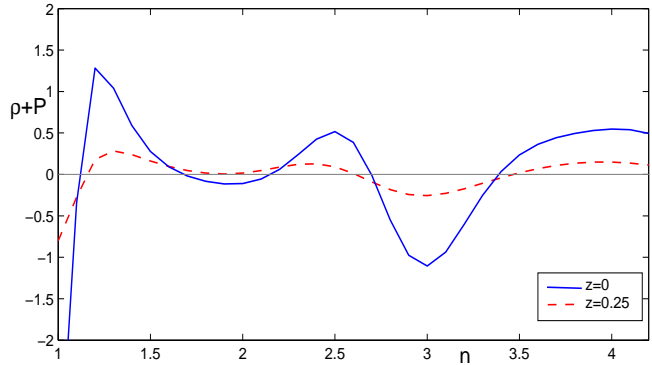} \vspace{4cm}
\end{center}
 \caption{\small {The null energy condition is respected in some subspaces
 of the model parameter space. }}
\end{figure}

\newpage
\section{Summary and Conclusion}
In this paper we have studied possible realization of the
phantom-like behavior in some viable $f(R)$ gravity models. By
phantom-like behavior, we mean increasing of the effective energy
density with cosmic time while the effective equation of state
parameter is less than $-1$. We have shown that some models of
$f(R)$ gravity can display a phantom-like behavior without violating
the null energy condition in some subspaces of their model parameter
space. To do this end, first we have considered a modified gravity
model with $f(R)=R+f_{0}R^{\alpha}$ and we found that the
phantom-like behavior can be obtained in the region of parameter
space with $0.5\leq\alpha<1$ and in this domain null energy
condition is respected. Although for negative values of $\alpha$ the
effective energy density has an increasing behavior with cosmic
time, but the null energy condition is violated. In the second
stage, we have considered a model of modified gravity with a $\ln R$
and an additional power law term. With a suitable choice of the
model parameters, this model which has potential to describe the
early time inflation and late time acceleration of the universe,
accounts for realization of the phantom-like behavior too. There are
appropriate subspaces of the model parameter space that null energy
condition is respected for this choice of $f(R)$. Finally we have
considered the Hu-Sawicki which has a very good phenomenology and
has been successful to pass the sever constraints imposed by solar
system tests and the perturbation theory. We showed that this model
is also capable to account for phantom like behavior without
violating the null energy condition with suitable choice of the
model parameters. The main feature of this work is the realization
of the phantom-like behavior without introducing any phantom fields
that violate the null energy condition in the spirit of modified
gravity.\\

{\bf Acknowledgement}\\
We would like to thank Prof. Sergei D. Odintsov for his invaluable
comments on this work.

\end{document}